\documentclass[%
 aip,
 jcp,
 amsmath,amssymb,
 reprint,longbibliography,
 citeautoscript, 
 floatfix,
 groupedaddress
]{revtex4-1}
\usepackage{graphicx}
\usepackage{dcolumn}
\usepackage{bm}

\usepackage[utf8]{inputenc}
\usepackage[T1]{fontenc}
\usepackage{mathptmx}
\usepackage{etoolbox}

\usepackage{setspace}
\usepackage{amsmath}
\usepackage{amssymb}
\usepackage{amsthm}
\usepackage{physics}
\usepackage{bbold}
\usepackage{bm}
\usepackage{hyperref}

\makeatletter
\def\@email#1#2{%
 \endgroup
 \patchcmd{\titleblock@produce}
  {\frontmatter@RRAPformat}
  {\frontmatter@RRAPformat{\produce@RRAP{*#1\href{mailto:#2}{#2}}}\frontmatter@RRAPformat}
  {}{}
}%
\makeatother
\begin{document}

\preprint{AIP/123-QED}

\title{Accurate Numerical Simulations of Open Quantum Systems Using Spectral Tensor Trains}
\author{Ryan T. Grimm}
\author{Joel D. Eaves}
\homepage{Joel.Eaves@colorado.edu}
\affiliation{Department of Chemistry, University of Colorado Boulder, Boulder, CO 80309, USA\looseness=-1} 
\date{\today}

\begin{abstract}
Decoherence between qubits is a major bottleneck in quantum computations. Decoherence results from intrinsic quantum and thermal fluctuations as well as noise in the external fields that perform the measurement and preparation processes. With prescribed colored noise spectra for intrinsic and extrinsic noise, we present a numerical method, Quantum Accelerated Stochastic Propagator Evaluation (Q-ASPEN), to solve the time-dependent noise-averaged reduced density matrix in the presence of intrinsic and extrinsic noise. Q-ASPEN is arbitrarily accurate and can be applied to provide estimates for the resources needed to error-correct quantum computations. We employ spectral tensor trains, which combine the advantages of tensor networks and pseudospectral methods, as a variational ansatz to the quantum relaxation problem and optimize the ansatz using methods typically used to train neural networks. The spectral tensor trains in Q-ASPEN make accurate calculations with tens of quantum levels feasible. We present benchmarks for Q-ASPEN on the spin-boson model in the presence of intrinsic noise and on a quantum chain of up to 32 sites in the presence of extrinsic noise. In our benchmark, the memory cost of Q-ASPEN scales linearly with the system size once the number of states is larger than the number of basis functions.
\end{abstract}

\maketitle
\section{Introduction}

An open quantum system undergoes irreversible relaxation, and describing its dynamics remains an open theoretical challenge in condensed matter. The typical scenario in an open quantum system is one where an external field prepares the system in a nonequilibrium state that decays to thermal equilibrium in the absence of that field \cite{weiss2012quantum}. This scenario---the initial state decay problem---represents many experiments in chemical physics, including condensed phase electronic spectroscopy, energy transfer, and electron transfer \cite{nitzan2006chemical}. 

However, the initial state decay problem does not cover all cases of condensed phase quantum relaxation, including those applications in quantum control \cite{Paz-Silva2014-vp} and quantum information \cite{Paz-Silva2017-oo} that have become increasingly relevant to chemical physics. In these applications, an external field does not disappear at time zero but instead keeps the system away from thermal equilibrium. 

The fields and their fluctuations can be too strong for perturbation theory. In electron and nuclear spin magnetic resonance, for example, the external Zeeman field forces the system into a nonequilibrium steady state by splitting otherwise degenerate M-sublevels. Unlike the quantum and thermal fluctuations that control relaxation in the initial state problem, fluctuations in the external fields do not need to be balanced by dissipation. The time-dependent fields that perform the gate operations in a quantum circuit, like ''$\pi$-pulses'', are another noise source. Like the Zeeman field, they are not subject to dissipative forces. 

Borrowing language from van Kampen \cite{van1992stochastic}, we call the fluctuations induced by external fields ``extrinsic noise'' and the thermal \& quantum fluctuations ``intrinsic noise.'' These two noise sources have fundamentally different properties. Unlike extrinsic noise, intrinsic noise must obey the fluctuation-dissipation theorem. Additionally, correlations can be quite different between extrinsic and intrinsic noise sources. It is a standard assumption that relaxation dynamics induced by thermal fluctuations are uncorrelated between different states of a quantum system. Field fluctuations, on the other hand, can induce correlated fluctuations between states. In magnetic resonance experiments on molecules, the wavelengths of the fields are long compared to interparticle spacings, and the fluctuations between quantum levels on different molecules are correlated. 

In a quantum information context,  intrinsic and extrinsic noises both lead to errors in quantum circuits \cite{Paz-Silva2017-oo}. Even so, quantum error correction can make a quantum computer viable \cite{Preskill2018-zi}. Indeed, the viability of a quantum computer may depend on the resources required to error-correct. It is, therefore, vital to have models that translate how various features of intrinsic and extrinsic noise sources lead to uncertainties in quantum calculations. This paper presents an accurate numerical method capable of modeling the relaxation of a system's reduced density matrix in the presence of intrinsic and extrinsic noise sources without introducing an auxiliary heat bath. 

An outstanding problem in numerical approaches to quantum noise is the ``curse of dimensionality,'' the exponential increase in the size of the Hilbert space with the number of qubits. For example, the smallest error-correcting quantum circuit requires five qubits \cite{Laflamme1996-gy}---a Hilbert space of $2^5 = 32$ dimensions. Current quantum processors have a massive state space \cite{Arute2019-mk, Google-Quantum-AI2023-ot}. Google's 53 qubit Sycamore processor, for instance, has a Hilbert space of $2^{53} \approx 10^{16}$ dimensions \cite{Arute2019-mk}. Due to decoherence, the system does not undergo purely unitary motion in the full Hilbert space. For many initial conditions and control sequences, the system evolves in a space of reduced dimension, making numerical simulations of the quantum circuit feasible on a classical computer. In practice, however, finding an efficient and accurate compression scheme of the full-dimensional system into a lower-dimensional one is difficult, to say the least. A clever ansatz for the relevant excited states --- like in the work of Chan \& Reichman \cite{Ng2023-zl, Park2024-fm} and Abanin \cite{Lerose2021-ru, Sonner2021-vn, Thoenniss2023-mj} --- can accomplish it for certain model problems, but a general approach remains elusive.  However, there has been significant progress made for some quantum impurity problems \cite{Ng2023-zl, Park2024-fm, Sonner2021-vn, Lerose2021-ru, Thoenniss2023-mj, Nunez_Fernandez2022-nj}.

The method presented here compresses the dynamical space with a generic ansatz for the time evolution of the density matrix using spectral tensor trains \cite{Bigoni2016-kn}, combining the advantages of pseudo-spectral methods and tensor compression to flatten the high-dimensional space accurately and efficiently. Once the number of states exceeds the number of basis functions for the spectral expansion of a relaxation operator, the method scales linearly with the number of states.

The structure of the stochastic Liouville equation (SLE) for the reduced density matrix in the quantum relaxation problem is isomorphic to the equation of motion for a system of classical oscillators with multiplicative noise. In a previous paper, we showed how to solve for the average dynamics in such a problem without averaging over stochastic trajectories. Our previous method \cite{Grimm2024-yg}, ASPEN, discretizes the time to an evenly spaced grid and finds an exact mapping between the resulting state-time space and a one-dimensional spin chain. Computing the noise-averaged trajectories is formally equivalent to calculating a partition sum over all trajectories on the spin chain. Except for Markovian noise, computing the sum is nontrivial because the spin-spin interactions can be long-ranged with a length scale set by the correlation function of the noise.  Calculating the partition sum incurs a computational cost that grows exponentially in the number of discrete time points $N$ required to represent the trajectory accurately. Because the quasi-adiabatic path integral (QUAPI) method \cite{Makarov1994-xb, Makri1995-gm, Makri1995-ns} has such a similar mathematical structure to ASPEN---particularly the mapping to the one-dimensional quantum chain---we made ASPEN tractable by borrowing the tensor network topology from time-evolving matrix product operator (TEMPO) method\cite{Strathearn2018-jc, Gribben2022-vg} used to accelerate the convergence of QUAPI. The result is an efficient and accurate numerical solution method based on a tensor network whose cost grows only linearly in the ``memory length'' $M$, which is related to the length scale of the spin-spin interactions. The remarkable scaling properties of tensor networks make numerically exact solutions tractable. In many cases --- such as the one-dimensional stochastic oscillator \cite{Grimm2024-yg} and the spin-boson model near its localization-delocalization phase transition \cite{Strathearn2018-jc} --- it is feasible to simulate the dynamics without a memory cutoff.  More generally, when the correlation decays sufficiently quickly, M is much less than N, and one can discard interactions beyond the memory length. 

In this paper, we revisit the quantum relaxation problem using the SLE for the reduced density matrix and illustrate a quantum version of ASPEN called Q-ASPEN. Q-ASPEN is a numerical method that solves the reduced density matrix of quantum states in the presence of intrinsic and extrinsic noise sources. Like ASPEN and QUAPI, we phrase the solution of the time-dependent reduced density matrix into the problem of calculating the partition sum of a one-dimensional quantum chain. 

A critical innovation in Q-ASPEN is the introduction of spectral tensor trains \cite{Bigoni2016-kn, Gorodetsky2019-ij} (STTs) to, for example, reduce a many-dimensional integral into many one-dimensional integrals. To our knowledge, STTs have seen limited attention in the chemical and condensed matter physics communities \cite{Greene2017-lx, Lyu2022-od, Lyu2023-oi} . In Q-ASPEN, we use an STT as a matrix product ansatz for a multi-dimensional function with upwards of 50 dimensions. In the same way a neural network can be viewed as a generalizable and efficient approximation to a many-dimensional function, so can the STT. We employ the same methods used to train neural networks to optimize the STT \cite{Gorodetsky2018-sf}. Using the STT, computing the partition sum becomes tractable because we systematically find the states of highest weight in the partition sum through tensor compression and spectral decomposition. In cases where Q-ASPEN and TEMPO are comparable, the results of the two methods are nearly indistinguishable. However, Q-ASPEN can scale linearly with the number of quantum states, so some numerical solutions are possible with Q-ASPEN when they are computationally out of reach for TEMPO. 

\section{Methods}
With units of time by energy chosen in terms of $\hbar$,  we begin the derivation of Q-ASPEN with the SLE for the reduced density matrix $\bm{\rho}(t)$ of the quantum system in the presence of an extrinsic stochastic field
\begin{equation}
    i\frac{d\bm{\rho}(t)}{dt} = \bm{{\cal L}}(t) \bm{\rho}(t),
    \label{eq:liouville}
\end{equation}
where $\bm{{\cal L}}(t) = \bm{L}_0 + \xi(t)\bm{L}_1$ is the time-dependent Liouvillian, $\xi(t)$ is the noise, ${\bm L}_0$ is the Liouvillian of the system in the absence of the noise, and ${\bm L}_1$ is the time-independent coupling matrix between the noise field and the system. In the examples here, ${\bm L}_0$ corresponds to an exactly solvable Hamiltonian $\bm{H}_0$ so that $\bm{L}_0 $ in Liouville space corresponds to $\bm{H}_0$ in Hilbert space via $\bm{L}_0 \bm A  \leftrightarrow [\bm{H}_0,\bm{A}]$. $\bm{V}$ is a time-independent operator in the system space that couples it to the noise where, similarly, $\bm{L}_1 \bm A  \leftrightarrow [\bm V,\bm{A}]$. For many external noise problems $\bm V$ is the Hamiltonian for the light-matter interaction. The noise field is a Gaussian random variable with zero mean, so its statistics are completely specified by its two-point time correlation functions. 

Introducing noise into a theory is a shortcut to statistically representing many degrees of freedom in a complex system. However, shortcuts can be dangerous, and simulating noisy quantum observables is subtle. Should one include the noise in the quantum time evolution equations of the system, or is it only rigorous to include the noise fields for the macroscopic averages after the quantum averages are done? Some clarifying comments about classical and quantum noise-averaging are in order. 

There are two equivalent approaches in classical probability theory for random variables under the influence of additive Markovian noise, like in Brownian motion. One can either consider stochastic trajectories of the random variable or solve for the time-dependent distribution function of the random variable directly. The Langevin equation describes the former, and the Fokker-Planck equation the latter. The Fokker-Planck equation for the probability density of the random variable has a dissipative term that arises from the fluctuations in the Langevin equation. Eq. \ref{eq:liouville} looks like a Langevin equation for a probability density, so it looks inconsistent a priori. In his work on the SLE \cite{Kubo1963-iz}, Kubo noted that Eq. \ref{eq:liouville} contains fluctuations without dissipation. As a result, detailed balance cannot be satisfied, and populations will relax to their infinite temperature values at asymptotically long times \cite{Moix2013-ch}. Thus, Eq. \ref{eq:liouville} may be valid for extrinsic noise but not intrinsic noise. Nonetheless, several examples of theoretical treatments ignore that fact, do not clearly distinguish between extrinsic and intrinsic noise, and live with the consequences. 

There are several ways \cite{Tanimura1989-pa, Stockburger2004-sg, Strunz1996-md} to restore detailed balance in the SLE for intrinsic noise. The most common approach is to choose a different starting point and write down the Liouvillian for a closed quantum system, introducing a heat bath that is linearly coupled to the system \cite{Feynman1963-ax}. Just as it does in the classical analog, the coupling term plays the role of the stochastic field and linear response ensures that the dynamics of the reduced density matrix with the bath traced out obey detailed balance. The penalty is that one must solve a complex time-dependent many-body problem for the reduced density matrix. This is a fruitful approach in problems like spectroscopy, polaron transport, energy transfer, and electron transfer, where the only relevant noise is intrinsic, and perturbation theory is usually good enough. In those problems, the bath corresponds to normal modes of real nuclear vibrations, and one can explicitly compute the spectral density of coupling constants between the electronic subsystem and the vibrations. However, introducing a heat bath is unphysical and distasteful when considering both intrinsic and extrinsic noise. Introducing spectral densities for the bath, or baths, that obey the Zeroth law of thermodynamics and simultaneously represent the statistics of intrinsic noise at finite temperature and external noise---presumably at infinite temperature---is not the approach taken here. 

Based on Tanimura's earlier work \cite{Tanimura1989-pa} with Kubo, Tamimura showed how to include the dissipative term in the SLE for the density matrix without the detour of introducing a fictitious bath \cite{Ishizaki2005-hq}. By expanding the dissipation operator of the random field in terms of its modes, one arrives at the hierarchical equations of motion (HEOM) for the reduced density matrix. The HEOM are a powerful workhorse of numerical quantum dynamics for multi-level systems coupled to intrinsic, non-Markovian noise sources \cite{Tanimura2014-ek, Tsuchimoto2015-hw, Tanimura1989-pa, Ishizaki2005-hq, Wilkins2015-vy}. Another approach due to Stockburger \cite{Stockburger1999-zs, Stockburger2004-sg} keeps the structure of Eq. \ref{eq:liouville} but with an anti-commutator term coupled to a different noise field. The two noise fields are complex-valued with cross-correlations given by detailed balance. In principle, it is possible to solve for the reduced density matrix to arbitrary numerical accuracy by trajectory-averaging Stockburger's SLE. In practice, the SLE can be very slow to converge \cite{Grimm2024-yg, Stockburger2004-sg}.

Our work uses Stockburger's principal results as a starting point. We do not address the more subtle question about how one should interpret individual noisy trajectories of the reduced density matrix. We only view the SLE as a vehicle to arrive at its noise-averaged trajectories, which are physically meaningful. Intrinsic noise is a complex-valued stochastic process, and extrinsic noise is a real-valued one. The former satisfies detailed balance, and the latter does not. When the noise is intrinsic, ${\bm L}_1$ acquires an anticommutator term multiplied by an additional noise field, $\xi(t){\bm L}_1 \rightarrow \xi(t){\bm L}_1^- + \nu(t){\bm L}_1^+$, where ${\bm L}_1^-$ corresponds to a commutator and ${\bm L}_1^+$ corresponds to an anticommutator in Hilbert space. Detailed balance relates $\xi(t)$ to $\nu(t)$. 

Rather than solve the SLE for trajectories and then average them, we directly find the noise-averaged reduced density matrix, generalizing the methods deployed in our earlier work on ASPEN \cite{Grimm2024-yg}. Our method, quantum-ASPEN (Q-ASPEN) forms an interesting historical closure, akin to Kubo's original work \cite{Kubo1963-iz} on the SLE from 1963, but for an SLE capable of modeling both intrinsic and extrinsic noise in quantum systems, faithfully including the fluctuation dissipation theorem for intrinsic noise.

The derivation here is very similar to ASPEN but differs somewhat in notation and technical detail. The notation gets cumbersome when several intrinsic and extrinsic noise fields are present. To keep the derivation legible and simple, we consider an SLE, Eq. \ref{eq:liouville}, with one extrinsic noise field $\xi(t)$. However, extensions to multiple noise fields with various auto- and cross-correlations are straightforward and the derivation for intrinsic noise appears in Appendix A. As the mathematicians say, ``without loss of generality'' nothing essential to the reader will be lost by considering the simple case of Eq. \ref{eq:liouville}. 

We assume prior knowledge of both intrinsic and extrinsic noise statistics, including their correlations with one another and with each other. These are usually given by experimental conditions where, for example, the intrinsic noise spectrum would describe the relaxation of the system in the absence of a field and the extrinsic noise spectrum would come from instrument responses that include the fields. In this case, intrinsic noise and extrinsic noise would be  uncorrelated with one another. As test cases, we present applications to simulating quantum dynamics in the presence of both uncorrelated intrinsic noise and correlated extrinsic noise. 

The formal solution to Eq. \ref{eq:liouville} for one realization of the noise field is a time-ordered exponential,
\begin{equation}
    \bm{\rho}(t) = \exp_{\leftarrow}\left[-i\int_{0}^t ds \bm{\mathcal{L}}(s)\right]\bm{\rho}(0).
\end{equation}
Using brackets $\langle \cdot \rangle$ to represent an average over the noise, the noise average over a system variable $\bm{A}$ accounts for quantum and noise fluctuations by $\langle A(t) \rangle  = \langle \Tr(\bm A \bm{\rho}(t))\rangle = \Tr(\bm A \langle \bm{\rho}(t)\rangle)$. The trace accounts for the quantum fluctuations---those fluctuations rooted in the uncertainty principle---that van Kampen originally termed ``intrinsic noise.'' The noise-averaged density matrix $\langle \bm{\rho}(t)\rangle$, therefore, describes the relevant state of the system. Averaging $\bm{\rho}(t)$ over all realizations of the noise gives the formally exact solution in terms of an initial value problem,
\begin{equation}
    \langle \bm \rho(t)\rangle = \bm{\Phi}(t) \bm{\rho}(0).
\end{equation}
Kubo's relaxation operator $\bm{\Phi}(t)$ plays the role of the Green's function in classical probability theory for the Fokker-Planck equation and gives the exact solution to the density matrix at time $t$ for an arbitrary initial condition. Except for the trivial cases where ${\cal L}(t)$ commutes with itself at all times, it is not easy to find an analytical solution to ${\bm \Phi}(t)$.

\begin{figure}
    \centering
    \includegraphics[keepaspectratio, width=8.6cm]{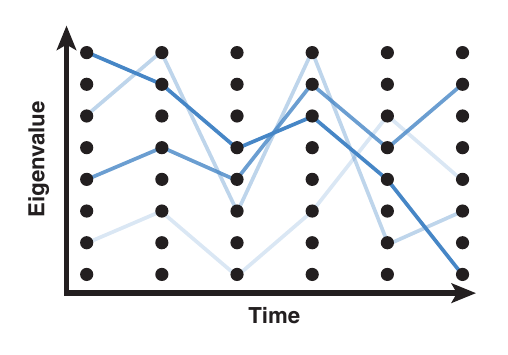}
    \caption{An illustration of a set of walks in the space of eigenfrequencies of $\bm L_1.$ We calculate the Q-ASPEN propagator from a weighted sum of these trajectories, which we illustrate as blue lines. Black dots represent the space-time points on the grid of eigenfrequencies of $\bm L_1$. The intensity of a pathway represents its weight. }
    \label{fig:traj}
\end{figure}

The objective of ASPEN and Q-ASPEN is to calculate $\bm{\Phi}(t)$ numerically by discretization,  Trotter splitting, and cumulant averaging. We discretize the trajectory into $N$ timesteps of length $\tau$ and let $t_k \equiv \tau k$. Using the composition property, we break the propagation into steps of length $\tau$
\begin{equation}
\bm{\Phi}_N = \left\langle \prod_{k=1}^N \exp_\leftarrow(-i \bm{L}_0 \tau - i\xi_k\bm{L}_1) \right\rangle,
\label{eq:thething}
\end{equation}
where $\bm{\Phi}(t_N) \equiv \bm{\Phi}_N,$ and $\xi_k = \int_{t_{k - 1}}^{t_k} ds \; \xi(s)$. In ASPEN, there is a defective matrix that one has to express as a sum of a Hermitian and anti-Hermitian matrix that do not commute with one another. The upshot is that we split the stochastic integration along a time slice into an upper and lower branch. That complication is absent in Q-ASPEN. Applying symmetric trotter splitting to Eq. \ref{eq:thething} yields
\begin{equation}
    \bm{\Phi}_N \approx \left\langle \prod_{k=1}^N e^{-i\bm{L}_0\tau/2} e^{-i\xi_k\bm{L}_1} e^{-i\bm{L}_0\tau/2} \right\rangle.
\end{equation}
Inserting a complete set of states at each time slice $\ket{\omega}$ such that ${\bm L}_1\ket{\omega} = \omega\ket{\omega}$,  defining $\bm{E}(\omega) \equiv \ket{\omega}\bra{\omega}$ and $\bm{\mathcal{G}}_0(\omega) \equiv e^{-i\bm{L}_0\tau/2}\bm{E}(\omega)e^{-i\bm{L}_0\tau/2}$, 
\begin{equation}
    \bm{\Phi}_N \approx \sum_{\omega_N} \cdots\sum_{\omega_1} \left(\prod_{\alpha=1}^N\bm{\mathcal{G}}_0(\omega_\alpha)\right) \left\langle e^{-i \sum_{k=1}^N \xi_k \omega_k} \right\rangle,
\label{eq:Q-ASPEN}
\end{equation}
where the sums go over all eigenfrequencies of $\bm{L}_1$, and the subscripts on the eigenfrequencies $\{\omega_k\}$ denote the time-slices. Note that the product preserves positive time-ordering $\prod_{\alpha=1}^N\bm{\mathcal{G}}_0(\omega_\alpha) = \bm{\mathcal{G}}_0(\omega_N)\cdots \bm{\mathcal{G}}_0(\omega_1)$.

Owing to the Gaussian statistics of $\xi(t)$, the function ${\cal K}(\omega_N,\cdots,\omega_1) = 
\left \langle e^{-i\sum_{k=1}^N \xi_k \omega_k} \right \rangle$, called the influence kernel, has an \emph{exact}  Gaussian form, 
\begin{equation}
    {\cal K}(\omega_N,\cdots,\omega_1) = e^{-\frac{1}{2}\sum_{k,l=1}^N \omega_k G_{k,l}\omega_l},
    \label{eq:influence_kernel}
\end{equation}
where $G_{k,l}$ is a correlation matrix with elements 
\begin{equation}
    G_{k,l} = \int_{t_{k - 1}}^{t_k} ds \int_{t_{l - 1}}^{t_l} ds' \langle \xi(s)\xi(s')\rangle.
    \label{eq:corr}
\end{equation} 
analogous to to a discretized lineshape function. For extrinsic noise $G_{k,l} = G(k-l)$ and $G_{k,l} = G_{l,k}$. The relationships for complex-valued intrinsic noise are more complicated and appear in Appendix A.

\subsection{A statistical mechanics of trajectories and a mapping to an ideal gas} 
As in ASPEN, the relaxation operator $\bm{\Phi}$ of Q-ASPEN adopts the form of a partition sum, where the matrix product $\prod_{\alpha=1}^N\bm{\mathcal{G}}_0(\omega_\alpha)$ is akin to the fugacity product of $N$ particles. The effective Boltzmann factor $e^{-\beta V}$ has a pair potential $\beta V = \frac{1}{2}\sum_{k,l=1}^N \omega_k G_{k,l} \omega_l$ that weights each trajectory over eigenvalues. The partition sum is a Boltzmann weighted average over trajectories, Fig. \ref{fig:traj}, where each trajectory is a walk in discrete time over the space of eigenfrequencies.

Calculating such a partition sum directly is intractable because the number of terms in the sum grows exponentially in $N$. Because the interactions in the effective potential are not limited to nearest neighbors,  we cannot apply the transfer matrix method directly. An exactly solvable partition sum would take the form of an ideal gas of effective, or dressed particles,
\begin{equation}
  \bm{\Phi}_N = \sum_{\omega_N}\cdots \sum_{\omega_1} \prod_{\alpha=1}^N \tilde{\cal \bm{G}}_0(\omega_\alpha) = \prod_{k=1}^N \bm{Z}_k
  \label{eq:GCE}
\end{equation}
with $\tilde{\cal \bm{G}}_0(\omega_\alpha)$ corresponding to the dressed states in the chain and $\bm{Z}_k$ to their fugacities. Eq. \ref{eq:GCE} has the form of a mean-field theory, but one which is exact and not approximate. Such mappings between unsolvable problems in an interacting system to an exactly solvable ideal but dressed reference system are commonplace in condensed matter theory, from quasiparticle theories to bosonization to density functional theory. Usually, these mappings are restricted to model problems where the mapping comes from some combination of intuition and serendipity. Here, we accomplish such a transform using spectral tensor trains (STT), not approximately, but to \emph{arbitrary} numerical precision. 

\begin{figure}
    \centering
    \includegraphics[keepaspectratio, width=8.6cm]{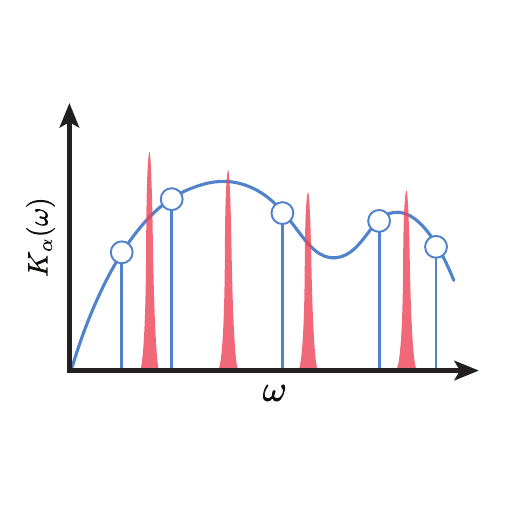}
    \caption{Illustrating the multidimensional integration method in Q-ASPEN, which converts an $N-$dimensional integral into $N$ one-dimensional integrals. A key step is to write a multidimensional function in the integrand as a spectral tensor train (STT), a product of matrix-valued continuous functions $\bm{K}_\alpha(\omega)$. The free propagator $\bm{\mathcal{G}}_0(\omega)$ is another matrix-valued function with discrete spectra (red) whose nonzero values do not necessarily lie on a quadrature grid. Expanding $\bm{K}_\alpha(\omega)$ in the basis of Chebyshev polynomials with variational matrix coefficients gives an optimal approximation (blue line) so that the target and fitted function match at the zeros of the Chebyshev polynomials (blue dots). The cost of Q-ASPEN scales as a fourth-order power law in the number of basis functions, which is usually much lower than the number of quantum states of the system. }
    \label{fig:interp}
\end{figure}

The first step is to make the ansatz that the influence kernel ${\cal K}(\omega_N,\cdots,\omega_1)$ can be written as a product of matrix-valued functions, generalizing the notion of the matrix product state in electronic structure theory. Decompose ${\cal K}$ as 
\begin{equation}
    {\cal K}(\omega_N,\cdots,\omega_1) = \bm{K}_N(\omega_N)\bm{K}_{N - 1}(\omega_{N - 1})\cdots \bm{K}_1(\omega_1)
    \label{eq:STT}
\end{equation}
where each $\bm K_\alpha$ has dimensions $B_{\alpha - 1} \times B_{\alpha}$. In the language of tensor networks, each $\bm{K}_\alpha$ is a spectral tensor core, or simply a core, and $\{B_{\alpha}\}$ are the bond dimensions. The edge cores $\bm{K}_1(\omega_1)$ and $\bm K_N(\omega_N)$ are row and column vectors, respectively, so that the product of matrices yields a scalar function. Substituting Eq. \ref{eq:STT} into Eq. \ref{eq:Q-ASPEN}, we get
\begin{equation}
    \bm{\Phi}_N \approx  \sum_{\omega_N}\cdots \sum_{\omega_1} \prod_{\alpha=1}^N\bm{\mathcal{G}}_0(\omega_\alpha) \otimes \left( \bm{K}_N(\omega_N) \cdots \bm{K}_1(\omega_1) \right)
\end{equation}
where inserting the tensor product $\otimes$ allows us to use the mixed product property $\bm{A}\bm{C} \otimes \bm{B}\bm{D} = (\bm{A}\otimes\bm{B})  (\bm{C}\otimes \bm{D})$ repeatedly, yielding
\begin{equation}
    \bm{\Phi}_N \approx  \prod_{\alpha=1}^N \sum_{\omega}\bm{\mathcal{G}}_0(\omega)\otimes \bm K_\alpha(\omega) 
\end{equation}
 Integrating over all eigenvalues, Fig. \ref{fig:interp}, gives the relaxation matrix as the partition function for an ideal gas, where the time-ordering makes the particles distinguishable
\begin{equation}
    \bm{\Phi}_N = \bm{Z}_N \bm{Z}_{N - 1} \cdots \bm{Z}_1
    \label{eq:Zs}
\end{equation}
with single particle partition functions $\bm{Z}_\alpha$ or fugacities
\begin{equation}
    \bm{Z}_\alpha \equiv \sum_\omega \;  \bm{\mathcal{G}}_0(\omega)\otimes \bm K_\alpha(\omega).
    \label{eq:part_func}
\end{equation}

\subsection{Extrinsic noise in the Markov limit} When the noise is extrinsic and Markovian, the partition sum is ideal. Let the Hamiltonian be $H = H_0 + \xi(t)V$. In the Markov case $\langle \xi(t) \xi(s) \rangle = \gamma \delta(t - s)$, where $\gamma$ is the noise strength. Then, the correlation matrix is $\bm G = \gamma \tau \mathbb{1}$, and $K_\alpha(\omega) = \exp(-\frac{\gamma \tau \omega^2}{2})$. The fugacities are identical and equal to
\begin{equation}
    \bm Z_\alpha = e^{-i\bm{L}_0\tau/2} e^{-\gamma \bm{L}_1^2 \tau/2} e^{-i\bm{L}_0\tau/2}.
\end{equation}
Applying the Baker–Campbell–Hausdorff formula and taking $\tau \to 0, N \to \infty$ with $t$ fixed in Eq. \ref{eq:Zs} yields an exact form for the relaxation operator 
\begin{equation}
    \bm \Phi(t) = \exp(-i t \bm L_0 -  t \gamma \bm L_1^2  / 2).
\end{equation}
After taking the time derivative, moving to Hilbert space, and expanding out $\bm L_1$ as an anti-commutator $\{A, B\} \equiv AB + BA$, the equation of motion is a Lindblad master equation \cite{Manzano2020-ci}
\begin{equation}
    \frac{d \langle \rho \rangle}{dt} = -i[ H_0,  \langle  \rho \rangle] + \gamma \left( V \langle  \rho \rangle  V - \frac{1}{2} \left \{ V^2, \langle \rho \rangle \right \} \right).
\end{equation}

\subsection{Decomposing the influence kernel using STT}

For non-Markovian fluctuations, we must find the STT decomposition of $\cal K$ numerically. We assert that the numerically exact non-Markovian system is Markovian in a higher dimensional space. The details of the derivation are in Appendix B. We generalize the transfer matrix trick that solves the one-dimensional Ising problems with nearest neighbor interactions to find the STT cores $\{ \bm{K}_n \}$. Write $\cal{K}$ as a product of transfer functions $\mathcal{T}_k$,  
\begin{equation}
     \mathcal{K}(\omega_N, \dots, \omega_1) =  \mathcal{T}_N(\omega_N, \dots, \omega_1) \otimes \cdots \otimes \mathcal{T}_1(\omega_1).
     \label{eq:K_transfer_function}
\end{equation}
While the product in Eq. \ref{eq:K_transfer_function} is a product of functions, inserting the tensor product between them makes representing them as matrix products notationally convenient. The $n$ arguments of each transfer function explicitly encode the memory effects between all $n$ points. Next, decompose each transfer function into another STT
\begin{equation}
    \mathcal{T}_n(\omega_n, \dots, \omega_1) = \bm \Gamma_{n \rightarrow n}(\omega_n) \cdots \bm \Gamma_{n \rightarrow 1}(\omega_1),
\end{equation}
where we call the cores   $\bm{\Gamma}_{n\rightarrow k}$  linking matrices. Using the mixed product property again (Appendix B), the relation 
\begin{equation}
       \bm K_n(\omega_n) = \bm \Gamma_{N \rightarrow n}(\omega_n) \otimes \cdots \otimes \bm \Gamma_{n \rightarrow n}(\omega_n)
       \label{eq:link_K}
\end{equation}
connects the linking matrices back to the cores of $\cal K$. Decomposing $\cal K$ into transfer functions allows us to make a finite memory approximation where $\mathcal{T}_n(\omega_n, \dots, \omega_1) \approx \mathcal{T}_{M + 1}(\omega_n, \dots, \omega_{n - M})$ for $n > M + 1$, where $M$ is the memory length. This is a major advantage as we only need to find the STT decomposition of functions with a dimension up to the $M + 1$ rather than the total number of time steps $N$.

We find the linking matrices by positing a parametric form for the cores and determining the parameters variationally. Motivated by the Weierstrass approximation theorem \cite{Trefethen2019-cl}, we expand the linking matrices in Chebyshev polynomials with matrix-valued coefficients, 
\begin{equation}
    \bm{\Gamma}_{n \rightarrow m}(\omega_m) \equiv \sum_k \bm{ A}^{(k)}_{n \rightarrow m} \psi_k(\omega_m),
\end{equation}
where $\psi_k$ is the $k$th Chebyshev polynomial, and $\{ \bm A^{(k)}_{n \rightarrow m} \}$ are the variationally determined coefficients. The number of basis functions is also a convergence parameter. In this work, we use the first ten Chebyshev polynomials. One could train the function anywhere in the spectral radius of $\bm{L}_1$, but it is most efficient to train them at the zeros of the Chebyshev polynomials \cite{Wu2017-rb}. This motivates the definition of the loss function $\Lambda(\bm{\theta})$
\begin{equation}
    \Lambda(\bm \theta) = \frac{1}{\cal N} \sum_{\bm \omega^*} \left( \mathcal{T}_n(\bm \omega^*) - \prod_{n = 1}^N \bm{\tilde{K}}_n(\omega_n^*)  \right)^2,
\end{equation}
where the parameter vector $\bm \theta$ consists of the entries of the coefficient matrices, and $\mathcal{N}$ is the number of training points. We compute the average loss over a set of training points on an $N$-dimensional grid of the zeros of the Chebyshev polynomials, where  $\bm \omega^* = (\omega^*_N, \dots, \omega^*_1)$  denotes a point on the grid. We randomly choose the points according to a prescription adapted from Wu and coworkers \cite{Wu2017-rb, Shin2017-vu},  who derive an optimally convergent importance sampling protocol to find tensor-product polynomial approximations to multi-variate functions. In short, we find the parameters of best fit by drawing random training points from their probability mass function on a Gauss-Chebyshev quadrature grid
and minimizing the loss function using stochastic gradient descent (SGD). We show the training curves in Fig. \ref{fig:train}. 

\subsection{Outline of the Algorithm}
The first step is to find the STT decomposition of each transfer function ${\cal T}_n$ up to the memory cutoff $M$ using the training procedure described above. We implement the training algorithm using the SGD and auto-differentiation functionality in \texttt{PyTorch} \cite{Paszke2019-cx}, which allows us to perform the training on the GPU. Next, for each eigenfrequency of $\bm{L}_1$, we use \texttt{TT-SVD} \cite{Oseledets2011-ni} to compress the free propagator $\bm{\mathcal{G}}_0(\omega)$ into a matrix product operator (MPO) to reduce the $\mathcal{O}(d^4)$ scaling of the free propagator. For every eigenfrequency, we construct an MPO representations of the summands in Eq. \ref{eq:part_func} by concatenating the MPO form of the free propagator with each of the linking matrices in Eq. \ref{eq:link_K}.  Taking the tensor product of an MPO with a matrix is exact because one simply appends a core with bond dimension one to the chain. We then sum over the MPOs using the method due to Oseledets \cite{Oseledets2011-ni}. We perform this procedure for each of the single-particle partition functions. We are left with an MPO representation of the product of single-particle partition functions in Eq. \ref{eq:Zs}. Finally, we contract the MPOs using White's \texttt{zip-up} algorithm \cite{Stoudenmire2010-ha}. 

To compress the free propagator, we first reshape  $\bm{\mathcal{G}}_0(\omega)$ from the $d^2 \times d^2$ matrix $\bm{\mathcal{G}}_0 (\omega; i,j)$ to the $4q$ rank $2 \times \cdots \times 2$ tensor $\bm{\mathcal{G}}_0(\omega; i_1, \dots, i_{2q}, j_1, \dots, j_{2q})$,  where $d$ is the dimension of the system, and $q = \log_2(d)$ is the number of qubits. Given the focus on quantum information, we assume a qubit base so that the system dimension is a power of two. Other choices of the base, like qudits, are possible, depending on the application.  We then decompose the tensorized version of the propagator into a matrix product operator (MPO) using the \texttt{TT-SVD} \cite{Oseledets2011-ci} algorithm to get $\bm{\mathcal{G}}_0(\omega; i_1, \dots, i_{2q}, j_1, \dots, j_{2q}) \approx \bm{U}_1(\omega; i_1, j_1) \bm{U}_2(\omega; i_2, j_2) \cdots \bm{U}_q(\omega; i_{2q}, j_{2q})$, where $\{\bm U_n\}$ are the cores of the MPO. \texttt{TT-SVD} uses truncated SVD to control the bond dimension, which discards singular values so that the relative error remains below  $\epsilon_{\text{SVD}}$. This procedure is similar to that of the time-evolving block decimation algorithm \cite{Vidal2003-ye} (TEBD), which simulates the action of a series of one and two-qubit gates on an MPS wavefunction. We, however, treat the propagator as a black box and decompose it algorithmically without reference to a quantum circuit.

\section{Results and Discussion}
\subsection{The Spin-Boson Model}

\begin{figure}
    \centering
    \includegraphics[keepaspectratio, width=8.6cm]{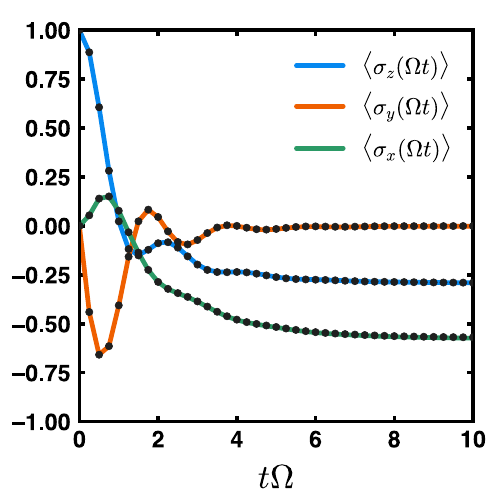}
    \caption{The dynamics of a biased, single qubit averaged over intrinsic noise whose spectrum matches an Ohmic spectral density. Q-ASPEN reproduces the results of the spin-boson model simulated using TEMPO. Choosing units in terms of the site tunnel coupling $\Omega$, the parameters are $\epsilon =1/ 2$, $\omega_c =1$, and $\beta = 1$. The coupling is $\alpha = 0.75$. The memory length, time step, and SVD cutoff are $M = 4$, $\tau = 0.25$, and $\epsilon_{\text{SVD}} = 10^{-8}$, respectively. All parameters are set identically for TEMPO, denoted by the black dots. The STT use the first $10$ Chebyshev polynomials as basis functions.  }  
    \label{fig:TLS}
\end{figure}

To benchmark Q-ASPEN, we simulate the dynamics of the spin-boson model and compare the results to those from TEMPO. In the language of open quantum systems, the spin-boson model consists of a two-level system coupled to a harmonic bath. In our language, the bath is replaced by intrinsic noise. The Hamiltonian for the system is $\bm H_0 = \Omega \bm  \sigma_x + \epsilon \bm \sigma_z$, where $\Omega$ is the site-site coupling, and $\epsilon$ is the energy gap. The fluctuations couple to the system through the operator $V = \alpha \bm \sigma_z$, where $\alpha$ is the coupling strength. We consider an Ohmic spectral density with an exponential cutoff of the form $J(\omega) = \omega \exp(-\omega / \omega_c),$ where $\omega_c$ is the cutoff frequency. The stochastic Liouvillian has real and imaginary noise fields that modulate the energy gap, whose correlation functions are entirely determined by $J(\omega)$ and the inverse temperature $\beta$ (Appendix A). Q-ASPEN is equivalent to computing the time-dependent density matrix for the initial-state relaxation problem in contact with a harmonic heat bath (see Appendix B). We report the time-dependent Pauli matrices $\bm \sigma_x, \bm \sigma_y, \bm \sigma_z$ because they, and the identity matrix, span the space for $\rho$ in the spin-boson problem. 

Fig. \ref{fig:TLS} shows that Q-ASPEN yields nearly identical dynamics to TEMPO for each of the spin expectation values. Any difference results from a subtle loss (on the order of $10^{-3}$) of trace conservation, which occurs because the spectral tensor train decomposition of the influence kernel does not have any mechanism to ensure trace preservation. The approximation can always be made more accurate, however, by increasing the number of basis functions, bond dimensions, or precision of the SVD cutoff. This comparison shows that introducing intrinsic noise fields can replicate the effect of a bath. Further, including a dissipative noise field restores detailed balance, resulting in Boltzmann-weighted populations at long times. 

\subsection{Quantum Site Model}

\begin{figure*}
    \centering
    \includegraphics[keepaspectratio, width=17.2cm]{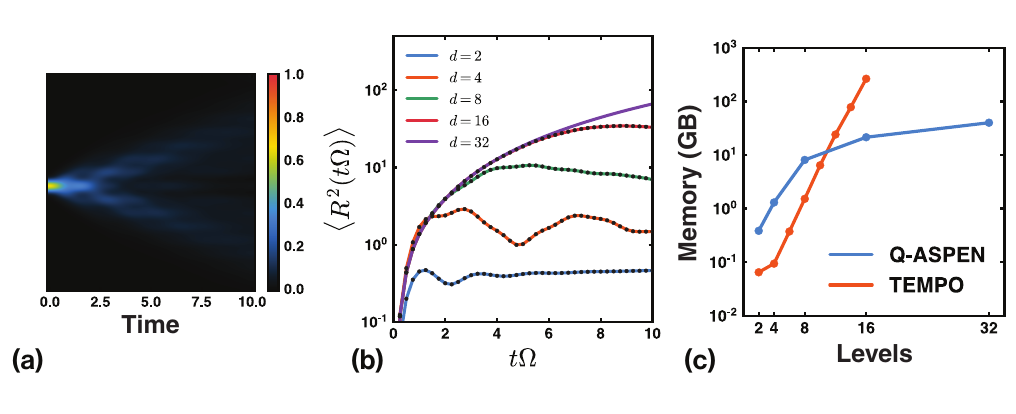}
    \caption{The dynamics of the one-dimensional quantum chain described in text for the one excitation subspace coupled to correlated, non-Markovian, extrinsic noise. Choosing units in terms of the intersite coupling $\Omega$ with $k_B = \hbar = 1$, the parameters $\beta = 1$, $\epsilon = 1$, and $\alpha = 0.5$. The convergence parameters are the same as in the spin-boson example. TEMPO results are black dots. Note that we omit the comparison to TEMPO for the thirty-two site case because it is not possible to compute it using TEMPO using our computational resources. To depict the transport dynamics of the excitation, we show (a) the population density as a function of time for the 32 site system and (b) mean squared displacement (MSD) as a function of time for each system size. The memory requirements for Q-ASPEN can be orders of magnitude less than those required for TEMPO.  } 
    \label{fig:ring_system}
\end{figure*}

To test Q-ASPEN on large quantum systems, we simulate the dynamics of a quantum chain model with $d$ sites. Each site couples to its nearest neighbor and experiences fluctuations from an extrinsic noise field. We consider periodic chains where $d$ ranges from 2 to 32, where each site has an alternating site energy and limit the state space to a single excitation basis. The system Hamiltonian in the site basis is $H_0 = \epsilon \sum_{n = 1}^d (-1)^{n} \ket{n}\bra{n} + \Omega  \sum_{n = 1}^{d}  \left(\ket{n} \bra{n + 1} + \ket{n + 1} \bra{n} \right),$ where $\ket{d + 1} = \ket{1}$, $\epsilon$ is the site energy, and $\Omega$ is the site-site coupling. The fluctuations modulate the energy of each site and couple to the system via $V = \alpha \sum_{n = 1}^{d} (-1)^n \ket{n} \bra{n}.$ In this example, $\bm L_0$ is the Liouville space representation of $H_0$ and $\xi(t) L_1 \rho \leftrightarrow \xi(t)[V,\rho]$. The sign of the coupling also alternates so that the noise on adjacent sites is anti-correlated. We use the same spectral density as in the spin-boson example to describe the noise, but we take the high-temperature limit of those fluctuations, in which case the noise becomes real-valued. The correlation function of the noise is related to the high-temperature limit of the Ohmic spectral density $\langle \xi(t)\xi(0)\rangle = \frac{1}{\beta} \int_0^\infty d \omega\,  J(\omega) \cos(\omega t) / \omega$ of the extrinsic noise \cite{nitzan2006chemical}.  We start the system with an excitation at the site at the origin and evolve the dynamics, Fig. \ref{fig:ring_system} (a), long enough (until $t \Omega = 10$) for the excitation to spread throughout the ring. We then calculate the mean squared displacement, Fig. \ref{fig:ring_system} (b), 
according to the expression $\langle R^2(t) \rangle = \sum_n n^2 \rho_{n,n}(t)$, where $n$ is the site index. The excitation encounters finite size effects before $t\Omega = 10$ except for the 32 site model, which undergoes ballistic transport $\langle R^2(t)\rangle \sim t^2$ until it saturates, presumably from finite size effects. Interestingly, there is no turn-over to a diffusive regime before saturation. 

To quantify the performance advantage of Q-ASPEN, we track the memory usage of both methods throughout the simulation and compute the average memory usage, Fig. \ref{fig:ring_system} (c). Q-ASPEN demonstrates significantly enhanced scaling over TEMPO.  The memory scaling shows two distinct regimes. TEMPO requires less memory for small systems because the number of basis functions needed to approximate the influence kernel exceeds the number of eigenfrequencies. However, as the system grows, the cost of approximating the influence kernel on a discrete grid scales aggressively \cite{Strathearn2018-jc} as $\mathcal{O}(d^8)$. In contrast, the cost of representing the influence kernel as a spectral tensor train is independent of system size. In Q-ASPEN, scaling with system size arises from two sources: the $\log_2(d)$ cores in free propagator and the bond dimension of the cores of the fugacities $\bm Z_\alpha$. The scaling of the latter is difficult to estimate as it is related to the compressibility of the $\bm Z_\alpha$. Because we calculate $\bm Z_\alpha$ from a sum of MPOs over the $d^2$ eigenfrequencies of $\bm L_1$, the bond-dimension scales as an unknown function of system size. Our empirical estimate,  Fig. \ref{fig:S1}, indicates that Q-ASPEN has \emph{linear} scaling with system size. However, a more rigorous analysis of the scaling is beyond the scope of this paper. Overall, this benchmark demonstrates the ability of Q-ASPEN to tackle very large systems without exorbitant computational resources.

\subsection{Training the STT: The Barren Plateau}
\begin{figure}
    \centering
    \includegraphics[keepaspectratio, width=8.6cm]{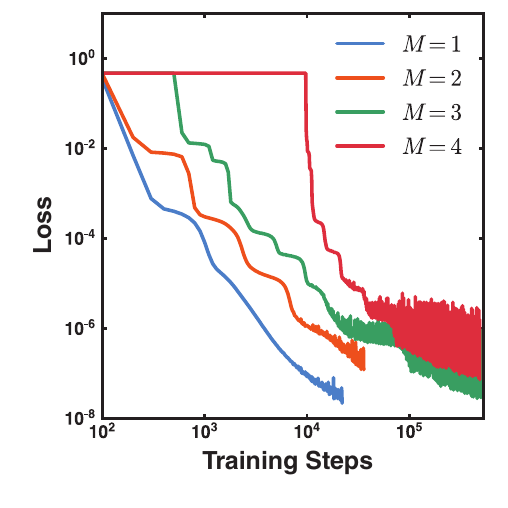}
    \caption{Training curves using stochastic gradient descent according to the importance sampling procedure described in the text for the spectral tensor train decomposition of the influence kernel $\cal{K}(\bm \omega)$ for different memory lengths from $M = 1$ to $M = 4$ for the spin-boson benchmark presented in Fig. \ref{fig:TLS}. For fixed bond dimensions, the loss function pauses for exponentially long times in the barren plateau region. }
    \label{fig:train}
\end{figure}

The main limitation of Q-ASPEN is in learning the STT representations of the transfer functions. Fig. \ref{fig:train} shows the training curve as a function of the number of gradient descent steps. Each of the training curves begins on a so-called "barren plateau."  The vanishing gradient problem, where the magnitude of the gradient vector vanishes as the number of parameters in the model increases, gives rise to the barren plateau \cite{Zhang2022-tc}. After leaving a barren plateau, however, the loss function decreases precipitously. Discrete tensor networks \cite{Liu2022-gr} and parameterized quantum circuits \cite{McClean2018-tk} are known to suffer from the barren plateau problem, where it is a major subject of research. Evidently, it also affects STTs as we observe longer barren plateaus with increasing memory length. 

\section{Conclusions}
In this paper, we introduced the Q-ASPEN method for simulating the dynamics of large quantum systems subject to realistic noise. We extended the numerically exact ASPEN method \cite{Grimm2024-yg} to solve the stochastic Liouville equation for the noise-averaged density matrix of an open quantum system. Importantly, our method satisfies detailed balance without introducing a fictitious heat bath and distinguishes between intrinsic and extrinsic noise sources. 

The Q-ASPEN propagator describes the dynamics of the reduced density matrix in its entirety. Calculating it is equivalent to computing the partition sum of a one-dimensional quantum chain, where the correlation function of the noise determines the length scale of the interactions between members of the chain. An important limit is the Markovian one, where the interaction length goes to zero and the propagator is exactly solvable because it reduces to a product of fugacity matrices---the partition function of an ideal gas. We developed a technique based on spectral tensor trains to factorize the influence kernel in the quantum relaxation problem, mapping a non-Markovian system to a Markovian one in higher dimensions. We use importance sampling in concert with methods popular in machine learning to find the optimal mapping from the interacting system to the ideal one. 

We demonstrated that Q-ASPEN can simulate the dynamics of a system in contact with a heat bath by considering the spin-boson model as a case of an intrinsic noise process. In this example, we showed that the two-level system relaxes towards Boltzmann equilibrium, as required by the fluctuation-dissipation theorem. The numerical results for the time-dependent density matrix in the initial-state relaxation problem are nearly indistinguishable from more conventional simulations with another method (TEMPO). Simulations of a quantum chain subject to extrinsic noise fluctuations also showed quantitative agreement with TEMPO, but it was only by using Q-ASPEN that we could simulate the system past 16 sites because of the large memory required in TEMPO. It is a computational chore to train the spectral tensor trains, but once they are trained, Q-ASPEN avoids a memory bottleneck. TEMPO scales as a steep power law $\mathcal{O}(d^8)$, where $d$ is the number of states, and Q-ASPEN scales \emph{linearly} for the one-dimensional quantum chain we tested. However, the barren plateau problem limits our ability to simulate systems with long-correlation times, implying that Q-ASPEN can be further accelerated using alternative training schemes. 

In Q-ASPEN, we tell the computer what it means to be an ideal gas and ask it to learn the optimal mapping from an interacting system to a noninteracting one. These mappings are central to many problems in condensed matter where one translates a strongly interacting system to a non-interacting one---particles to quasiparticles, particle-hole excitations to bosons, and so on---and we suspect that spectral tensor trains will have uses outside the application presented here. 

\begin{acknowledgments}
\vspace{-8pt}
R.T.G. was supported by the National Science Foundation Graduate Research Fellowship. This material is based upon work supported by the National Science Foundation Graduate Research Fellowship Program under Grant No. (DGE 2040434). J.D.E.'s work on the machine learning aspects of the manuscript was supported by the Department of Energy, Office of Basic Energy Sciences, under grant no. DE-SC0021387. Any opinions, findings, and conclusions or recommendations expressed in this material are those of the author(s) and do not necessarily reflect the views of the National Science Foundation.  This work utilized the Alpine high performance computing resource at the University of Colorado Boulder. Alpine is jointly funded by the University of Colorado Boulder, the University of Colorado Anschutz, and Colorado State University.
\end{acknowledgments}

\section*{Author Declarations}
\subsection*{Conflict of Interest}
The authors have no conflicts to disclose.
\subsection*{Author Contributions}
\textbf{Joel Eaves}: Review and editing (equal), Supervision (lead), Conceptualization (equal), Methodology (equal), Validation (equal).
\textbf{Ryan Grimm}: Review and editing (equal), Conceptualization (equal), Methodology (equal), Software (lead), Validation (equal).
\section*{Data Availability}
{\raggedright The data that support the findings of this study are available
from the corresponding author upon reasonable request. }

\appendix

\setcounter{section}{0}
\setcounter{equation}{0}
\setcounter{figure}{0}
\setcounter{table}{0}

\renewcommand{\theequation}{S\arabic{equation}}
\renewcommand{\thefigure}{S\arabic{figure}}

\section{Q-ASPEN with intrinsic noise}
We simulate the dynamics of the two-level system coupled to intrinsic noise using expressions from Stockburger \cite{Stockburger2004-sg}, who showed that the Feynman-Vernon influence functional of a linear response heat-bath Hamiltonian is exactly representable as a stochastic differential equation for the density matrix. The time-dependent Liouvillian is 
\begin{equation}
    \bm{\mathcal{L}}(t) \equiv \bm L_0 + \xi(t) \bm L_1^- + \nu(t) \bm L_1^+,
\end{equation}
where ${\bm L}_1^-$ corresponds to a commutator in Hilbert space and $\bm{L}_1^+$ to an anti-commutator. $\xi(t)$ and $\nu(t)$ are related by detailed balance. The correlation matrix of the noise fields is
\begin{align}
    \begin{bmatrix}
\langle \xi(t) \xi(t') \rangle & \langle \xi(t) \nu(t') \rangle \\
\langle \nu(t) \xi(t') \rangle & \langle \nu(t) \nu(t') \rangle \\
\end{bmatrix} &= \\ \begin{bmatrix}
S'(t - t') & 2i\theta(t - t')S''(t - t') \\
0 & 0
\end{bmatrix},
\end{align}
where the correlation function $S(t) = S'(t) + i S''(t)$ is
\begin{equation}
    S(t) \equiv \int_0^\infty d\omega \, J(\omega) \left[\cos(\omega t) \coth(\frac{\beta \omega}{2}) - i \sin(\omega t) \right],
\end{equation}
and $J(\omega)$ is the spectral density. Note how new symmetries emerge for the intrinsic case compared to the extrinsic case---the matrix elements of the correlation matrix transform differently under time reversal. The $\langle \xi(t) \xi(t') \rangle$ correlator is still even in time. However, the $\langle \xi(t) \nu(t') \rangle$ correlator is causal. The field $\nu(t)$ behaves as a response field to the fluctuating field $\xi(t)$ in the path-integral representation of a linear response theory. 

We apply symmetric Trotter splitting as above to get 
\begin{equation}
    \Phi_N  \approx \left \langle \prod_{n = 1}^N e^{-i \bm L_0 \tau/2} e^{-i \left[ \xi_k \bm L_1^- + \nu_k \bm L_1^+ \right]} e^{-i \bm L_0 \tau/2} \right \rangle.
\end{equation} Because $\bm L_1^+$ and $\bm L_1^-$ commute, we still only need to insert one resolution of the identity. However, $\bm L_1^+$ and $\bm L_1^-$ have different eigenvalues, so we must sum over the eigenvalues in pairs $(\omega_k^-, \omega_k^+)$. Thus, we obtain
\begin{widetext}
    \begin{equation}
\Phi_N  \approx \left \langle \prod_{k = 1}^N \exp(-i \bm L_0 \tau/2) \left[ \sum_{(\omega_k^-, \omega_k^+)} \bm E(\omega_k^-, \omega_k^+) \exp(-i \xi_k \omega^-_k  + \nu_k \omega^+_k) \right] \exp(-i \bm L_0 \tau/2) \right \rangle.
\end{equation}
\end{widetext}
Pulling the sums outside and grouping the scalar stochastic part yields
\begin{widetext}
        \begin{equation}
    \bm \Phi_N  \approx \sum_{(\omega_N^-, \omega_N^+)} \cdots \sum_{(\omega_1^-, \omega_1^+)}  \left[  \prod_{\alpha = 1}^N \bm{\mathcal{G}}_0(\omega_\alpha^-, \omega_\alpha^+)  \right] {\cal K}(\omega_N^-, \omega_N^+, \dots, \omega_1^-, \omega_1^+),
\end{equation}
\end{widetext}
where the function $\cal K$ has the form
\begin{equation}
    {\cal K}(\bm \omega^-, \bm \omega^+) \equiv \exp\left(-\frac{1}{2} \begin{bmatrix}
        \bm \omega^-  \bm \omega^+
    \end{bmatrix}
    \bm G
    \begin{bmatrix}
        \bm \omega^- \\ \bm \omega^+
    \end{bmatrix}
    \right), 
\end{equation}
where the correlation matrix is 
\begin{equation}
        \bm G = \begin{bmatrix}
        \langle \bm \xi \otimes \bm \xi \rangle & \langle \bm \xi \otimes \bm \nu \rangle \\ 
        \langle \bm \nu \otimes \bm \xi \rangle & \langle \bm \nu \otimes \bm \nu \rangle \\ 
    \end{bmatrix}.
\end{equation}

    From here, it should be clear how to generalize Q-ASPEN to more complex situations that include many noise sources. For example, if we were to add an extrinsic source of noise $\eta(t)$, the correlation matrix would be 
    \begin{equation}
        \bm G = \begin{bmatrix}
        \langle \bm \eta \otimes \bm \eta \rangle & \bm 0 & \bm 0 \\
        \bm 0 & \langle \bm \xi \otimes \bm \xi \rangle & \langle \bm \xi \otimes \bm \nu \rangle \\ 
        \bm 0 & \langle \bm \nu \otimes \bm \xi \rangle & \langle \bm \nu \otimes \bm \nu \rangle \\ 
    \end{bmatrix}.
    \end{equation}

\section{Generalization of the Transfer Matrix Method}
Although the spectral tensor train ansatz in Eq. \ref{eq:STT} is concise, computing the cores $\{\bm K_n\}$ directly is inefficient. Doing so means that for each time step, the decomposition needs to be recomputed, resulting in a growing computational cost with time. Thus, we decompose Eq. \ref{eq:influence_kernel} into transfer functions and compute the STT decomposition for each transfer function. It turns out that one only needs to decompose the transfer functions up to the memory length $M$. We show the derivation for an SLE with a single noise variable. However, the process generalizes trivially to multiple variables by interlacing the frequencies so that frequencies that are at the same time are adjacent in the parameter order: $\mathcal{K}((\omega_N^-, \omega_N^+), \dots, (\omega_1^-, \omega_1^+))$.

We first make a transformation such that the argument to the exponent is causal 
\begin{equation}
    \mathcal{K}(\omega_N, \dots, \omega_1) = \exp(-\sum_{n = 1}^N \sum_{\Delta = 0}^{n-1} \omega_n \tilde{G}_{\Delta} \omega_{n - \Delta}),
\end{equation}
where is the symmetrized correlation matrix \cite{Grimm2024-yg}. Breaking the product into transfer functions yields 
\begin{equation}
    \mathcal{K}(\omega_N, \dots, \omega_1) = \bigotimes_{n = 1}^N \mathcal{T}_n(\omega_n, \dots, \omega_1).
    \label{eq:K_ten_prod}
\end{equation}
Note that the tensor product and regular product are equivalent if the terms under the product are scalars. We define the transfer functions as
\begin{equation}
    \mathcal{T}_n(\omega_n, \dots, \omega_1) \equiv \exp(-\omega_n \sum_{\delta = 0}^{n - 1} \tilde{G}_\delta \omega_{n - \delta}),
    \label{eq:tranfer_decomp}
\end{equation}
where we have rewritten the casual expression in terms of $\delta \equiv n - m$, which encodes how far we are looking into the past. Working in terms of transfer functions allows us to make a very useful approximation that reduces dimensionality by throwing away arguments that correspond to times beyond the correlation time (memory length) of the system. More precisely $\mathcal{T}_n(\omega_n, \dots, \omega_1) \approx \mathcal{T}_{M + 1}(\omega_n, \dots, \omega_{n - M})$ provided that $\Tilde{G}_{M + 1} \approx 0$, where $M$ is the memory length. 

Instead of finding the STT of $\mathcal{K}$, we find the STT decomposition of each transfer function up to the memory length $M$. We then construct the cores of $\cal K$ from the cores of the transfer functions, which we call linking matrices. Write the $n$th transfer function as an STT to get
\begin{equation}
    {\cal T}_n(\omega_n, \dots, \omega_1) = \prod_{m = 1}^N \bm \Gamma_{n \rightarrow m}(\omega_m),
    \label{eq:linking_decomp}
\end{equation} where $\bm \Gamma_{n \rightarrow m} \equiv \mathbb{1}$ for $m > n$. The linking matrix $\bm \Gamma_{n \rightarrow m}$ links time $t = n\tau$ to  $t = m \tau$. 

Substitute Eq. \ref{eq:linking_decomp} into Eq. \ref{eq:K_ten_prod} and exchange the products using the mixed product property to get
\begin{equation}
    \mathcal{K}(\omega_N, \dots, \omega_1) = \prod_{n = 1}^N \bigotimes_{m = 1}^N \bm \Gamma_{m \rightarrow n}(\omega_n).
    \label{eq:K_link}
\end{equation}
Comparing Eq. \ref{eq:STT} to Eq. \ref{eq:K_link} yields a form for the cores of $\cal K$
\begin{equation}
    \bm K_n(\omega_n) = \bigotimes_{m = 1}^N \bm \Gamma_{m \rightarrow n}(\omega_n).
\end{equation}
Running the tensor product from $1$ to $N$ is necessary for each $\{ \bm K_n \}$ to have the right shape to contract with one another. However, for the purposes of illustration, we can drop the identity-valued linking matrices to get the more intuitive expression 
\begin{equation}
    \bm K_n(\omega_n) \sim \bigotimes_{m = n}^N \bm \Gamma_{m \rightarrow n}(\omega_n),
\end{equation}
where $\sim$ denotes similar to. In the main text, we do not make the distinction for the sake of clarity. 
\begin{figure}
    \centering
    \includegraphics[keepaspectratio, width=8.6cm]{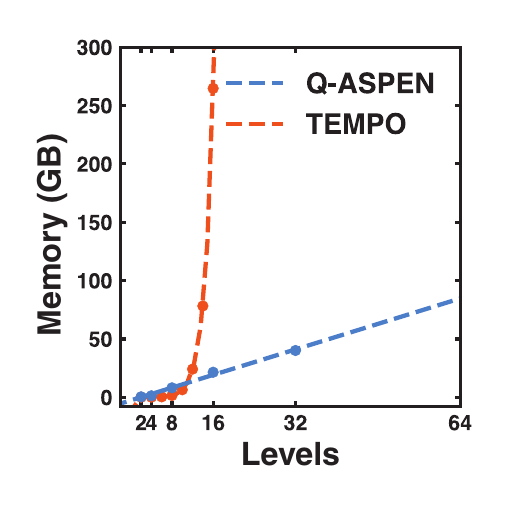}
    \caption{An alternate view of the memory scaling comparison of TEMPO and Q-ASPEN for the site benchmark, Fig. \ref{fig:ring_system}. We fit a linear polynomial ($R^2 = 0.993$) to the Q-ASPEN result and an eighth-order polynomial to the TEMPO result.}
    \label{fig:S1}
\end{figure}

\onecolumngrid
\vfill
\pagebreak
\twocolumngrid

\section*{References}
\vspace{-10pt}
\bibliography{references}

\end{document}